\documentclass[11pt]{article}
\usepackage{Moriond,epsfig}

\def\Journal#1#2#3#4{{#1} {\bf #2}, #3 (#4)}

\def\NPB{{\em Nucl. Phys.} B}
\def\PLB{{\em Phys. Lett.}  B}
\def\PRL{\em Phys. Rev. Lett.}
\def\PRD{{\em Phys. Rev.} D}

\def\PRV{{\em Phys. Rev.}}
\def\MPL{\em Mod. Phys. Lett.}

\def\be{\begin{equation}}
\def\ee{\end{equation}}
\def\bea{\begin{eqnarray}}
\def\eea{\end{eqnarray}}

\begin{document}

\hfill BUHEP-00-7 

\hfill hep-ph/0004238
\vspace*{3.5cm}
\title{STRONGLY-INTERACTING HEAVY FLAVORS \\ BEYOND THE STANDARD MODEL \footnote{Talk given at XXXVth  Rencontres de Moriond, 3-20-2000, Les Arcs, France.}}

\author{E.H. SIMMONS \footnote{e-mail: simmons@bu.edu}}

\address{Department of Physics, Boston University, 590 Commonwealth Ave.,\\
Boston, MA, 02215, USA}

\maketitle\abstracts{ The origin of mass must lie in physics beyond
the Standard Model.  Dynamical electroweak symmetry breaking models
like technicolor can generate masses for the $W$ and $Z$ bosons.
Providing the large top quark mass and large top-bottom mass splitting
while keeping $\Delta\rho$ and flavor-changing neutral currents small
requires new strong dynamics for the top and bottom quarks.  In
consequence, new particles are predicted at scales up to 10 TeV with
signatures in jets or heavy flavors. Searches for these states are
underway at Fermilab and LEP II.}

\section{Introduction}

The cause of electroweak symmetry breaking and the origin of fermion
masses remain central concerns of particle theory.  The Standard
Model, based on the gauge group $SU(3)_c \times SU(2)_W \times U(1)_Y$
accommodates fermion and weak boson masses by including a fundamental
weak doublet of scalar Higgs bosons ${\phi = {\phi^+ \choose \phi^0}}$
with potential function $V(\phi) = \lambda \left({\phi^\dagger \phi -
\frac12 v^2}\right)^2$.  However the Standard Model does not explain
the dynamics responsible for the generation of mass.

Furthermore, the scalar sector suffers from
\begin{figure}[b]
\centerline{\hbox{\epsfig{figure=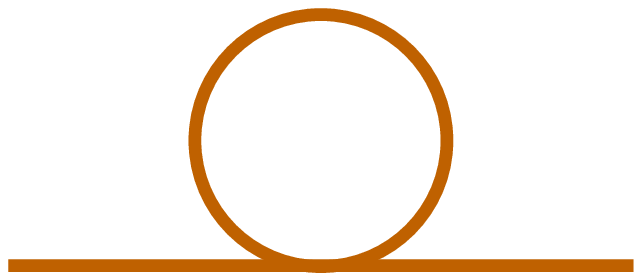,height=.35in}} \hspace{3cm} \hbox{\epsfig{figure=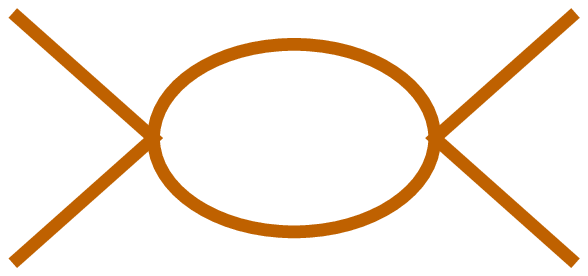,height=.35in}}}
\caption{At left, the gauge hierarchy problem: $ M_\phi^2\ \propto\ \Lambda^2$.  At right, triviality: $\beta(\lambda)\ = \ {{3\lambda^2}\over{2\pi^2}}\ > \ 0$.
\label{fig:hier}}
\end{figure}
two serious problems.  The scalar mass is unnaturally sensitive to the
presence of physics at any higher scale $\Lambda$ (e.g. the Planck
scale), as sketched in figure \ref{fig:hier}.  This is known as the
gauge hierarchy problem.  In addition, if the scalar must provide a
good description of physics up to arbitrarily high scale (i.e., be
fundamental), the scalar's self-coupling ($\lambda$) is driven to zero
at finite energy scales as indicated in figure \ref{fig:hier}.  But if
the scalar field theory is free (or ``trivial''), the electroweak
symmetry will not spontaneously break!  The scalars involved in
electroweak symmetry breaking must therefore be composite at some
finite energy scale. The Standard Model is only a low-energy
effective field theory and the origin of mass lies in physics outside
the Standard Model.

\section{Dynamical Symmetry Breaking}

In dynamical symmetry breaking theories, the compositeness of the
 scalar states becomes manifest at scales just above the electroweak
 scale $v \sim 250$ GeV.  In one realization called technicolor\,\cite{ehs-technicolor}, a new
 strong gauge interaction with $\beta < 0$ breaks the chiral
 symmetries of a set of massless fermions $T$ at a scale $\Lambda \sim
 1$ TeV.  If the fermions carry appropriate electroweak quantum
 numbers, the resulting condensate $\langle \bar T_L T_R \rangle \neq
 0$ breaks the electroweak symmetry.  The logarithmic
 running of the strong gauge coupling renders the low value of the
 electroweak scale (i.e.  the gauge hierarchy) natural.  The lack 
 of fundamental scalar bosons obviates concerns about triviality.

The quarks and leptons must couple to the source of electroweak
symmetry breaking, $\langle T\bar{T} \rangle$, in order to obtain
mass.  One possibility is to embed technicolor in a larger extended
technicolor (ETC) gauge group\,\cite{ehs-etc} under which all fermions
are charged.  When the ETC group breaks to its technicolor subgroup at
a scale $M_{ETC} > \Lambda_{TC}$, the ETC gauge bosons coupling
ordinary fermions $f$ to technifermions become massive. ETC
boson exchange then provides a contact interaction
\begin{equation}
\frac{g_{ETC}^2}{M_{ETC}^2} \bar{f_L}f_R \bar{T_L}T_R
\end{equation}
which yields a fermion mass $m_f \sim (g_{ETC}/M_{ETC})^2 (4\pi v^3)$ when the technifermions condense.

\begin{figure}
\vspace{-.5cm}
\hspace{-.7cm} \scalebox{.47}{\includegraphics{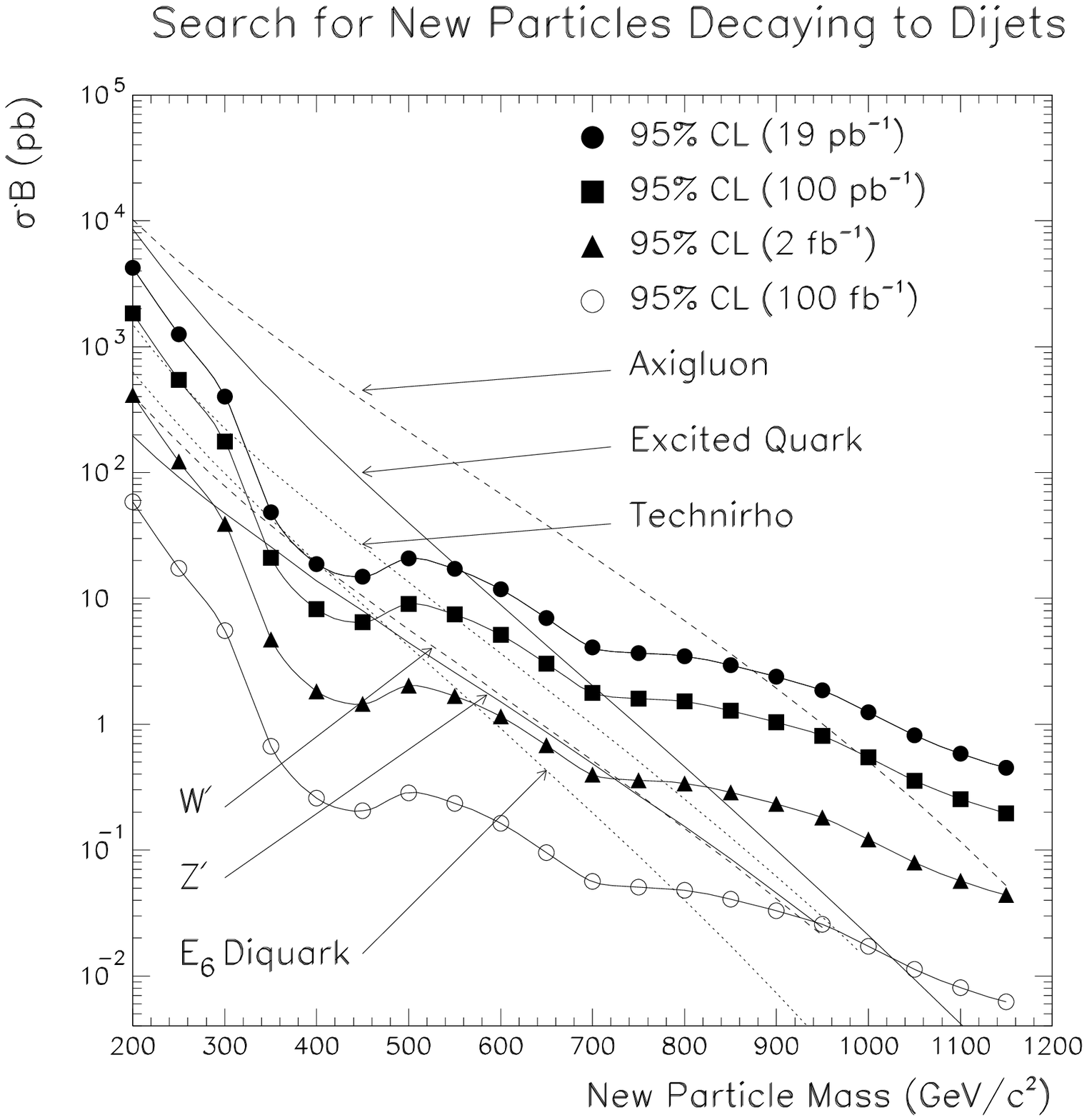}}
\hspace{4cm} \raise15pt\hbox{\scalebox{.5}{\includegraphics{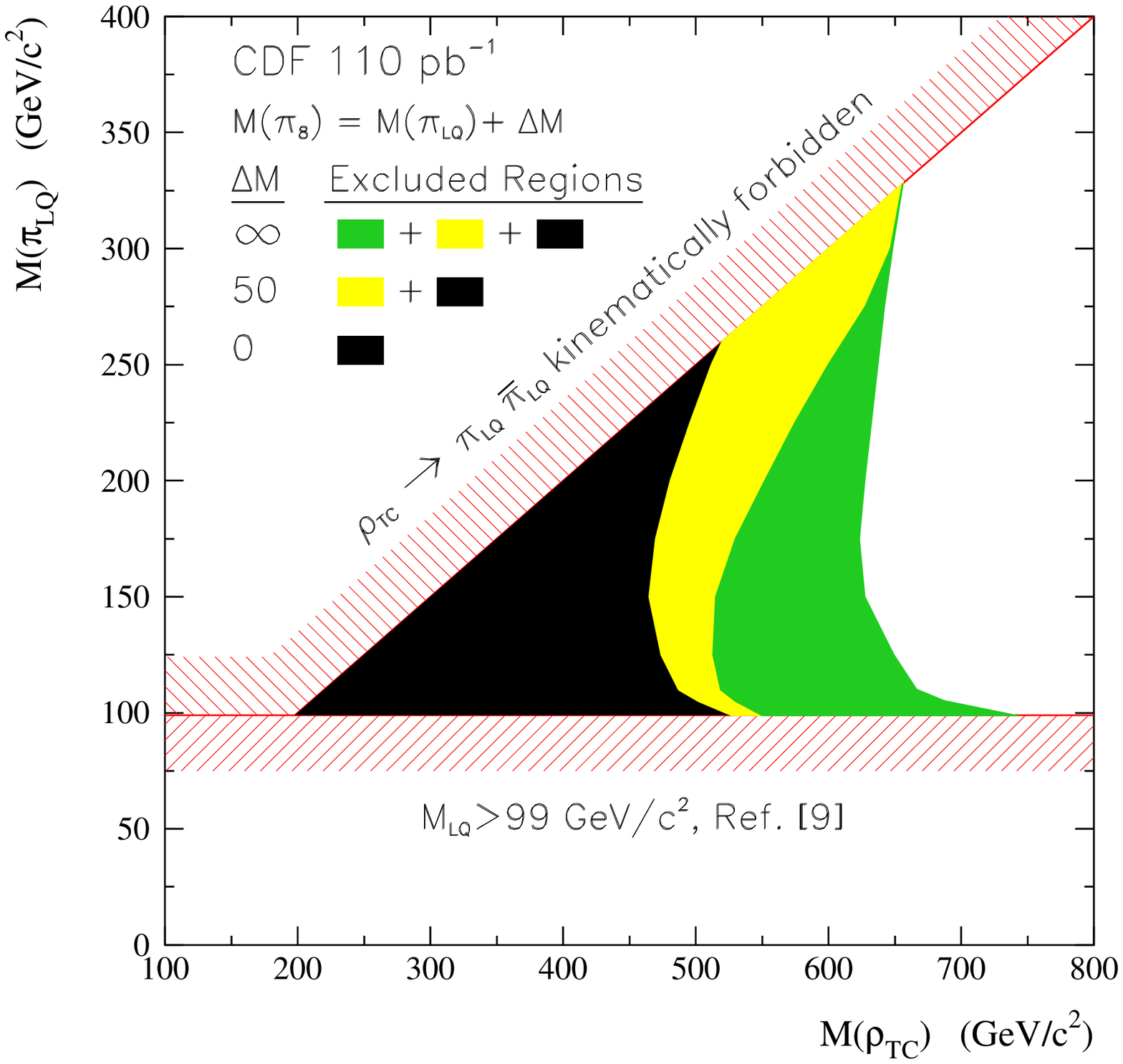}}}
\vspace{2cm}
\caption[cap]{Limits on technihadrons. Left: Projected limits\protect{\,\cite{ehs-tev2000}}on a color-octet technirho from $\rho_T \to jj$ at FNAL.  Right: CDF limit\protect{\,\cite{ehs-leptoq}} on $\rho_T$ and $\pi_{LQ}$ masses assuming the leptoquark technipion decays only as $\pi_{LQ} \to \tau b$}\label{fig:dual}
\end{figure}

Models of dynamical symmetry breaking include many technihadrons for
which experiments can search.  Their mass scale is set by the
electroweak scale, and their role in mass generation requires them to
couple to the Standard fermions and electroweak bosons.  Recent
searches\,\cite{ehs-pdg} for $\pi_T$, $\omega_T$ and $\rho_T$ include
many channels $\rho_T \to jj$; $\omega_T \to \gamma \pi_T$; $\rho_T
\to WW,\, W\pi_T,\, \pi_T \pi_T,\,\pi_{LQ}\pi_{LQ}$ [with $\pi_{LQ}
\to \tau q,\, \nu q$; $\pi_T \to b\bar b,\, c\bar c$] whose final
states involve jets or heavy flavors. Figure \ref{fig:dual}
illustrates the limits that FNAL searches for resonances decaying to
dijets can set on a color-octet technirho\,\cite{ehs-tev2000}.
Likewise, searches for leptoquarks\,\cite{ehs-leptoq} constrain a
$\rho_T$ decaying to a pair of technipions ($\pi_{LQ}$) with
third-generation leptoquark quantum numbers (figure \ref{fig:dual}).

These models also confront several key challenges.  The large value of
$m_t$ requires a small value for $M$ -- yet data on 
$K^0\bar{K}^0$ mixing imply\,\cite{ehs-etc} $M >$ 1000 TeV.
Sufficient weak isospin violation must be provided to split $m_t$ from
$m_b$ -- yet weak isospin breaking in technifermion couplings could
make $\Delta\rho$ too large\cite{ehs-appelquist}.  New strong dynamics
for the third family fermions can provide a solution.  Suppose that
extended technicolor provides only a small mass (and splitting) to the
top and bottom quarks and the technifermions at a high scale
$M_{ETC}$.  If new dynamics affecting only $t$ and $b$ turns on
at a scale $M$ intermediate between $M_{ETC}$ and $\Lambda_{TC}$, this
can provide a large $m_t$ and $m_t >> m_b$ without enlarging the
technifermion condensate or mass splittings.

\section{New Strong Top Dynamics}

New strong top quark dynamics is required to provide $m_t$ in a manner
consistent with existing data.  This dynamics forms a top quark
condensate that will contribute to (cf. topcolor\,\cite{ehs-topcolor})
or even dominate (cf. top mode\,\cite{ehs-topmode} or top seesaw\,\cite{ehs-topseesaw}) electroweak symmetry breaking.

Suppose the top quark feels a color interaction different from that
affecting the other up-type quarks\,\cite{ehs-topcolor}.  At high
energies, the strong gauge interactions would include both an
$SU(3)_h$ for the $t$ (and $b$) and an $SU(3)_\ell$ for the other
quarks.  To be consistent with existing hadronic data, these groups
spontaneously break to their diagonal subgroup (identified with
$SU(3)_{QCD}$) at a scale $M$: $SU(3)_h \times SU(3)_\ell \to
SU(3)_{QCD}$.  Thus, a color octet of heavy gauge bosons primarily
coupled to $t$ and $b$ exists at scales below $M$.  Their exchange
yields a new four-fermion interaction
\begin{equation}
- {4\pi\kappa \over M^2}
\left(\overline{t}\gamma_\mu {\lambda^a\over 2} t\right)^2
\end{equation}
that causes cause top quark condensation\,\cite{ehs-topmode} ($ \langle \bar{t}t\rangle \neq 0$) if the coefficient $\kappa$ is large enough.

Since these new strong interactions treat the top and bottom quarks
identically, one needs a mechanism to prevent the bottom quark from
condensing and becoming as massive as the top.  One model that
accomplishes this goal is the original topcolor-assisted technicolor\,\cite{ehs-topcolor}, which has the following gauge group and
symmetry-breaking pattern.
\begin{eqnarray}
G_{TC} \times SU(2)_W \times 
{U(1)_h \times U(1)_\ell} &\times& {SU(3)_h \times SU(3)_\ell}\nonumber  \\
&\downarrow& \ \ M \geq 1 {\rm TeV} \nonumber \\
G_{TC} \times SU(2)_{W} &\times&
  {U(1)_Y} \times {SU(3)_C} \nonumber \\
&\downarrow&\ \ \ \Lambda_{TC}\sim 1 {\rm TeV} \nonumber \\
{U(1)_{EM}} &\times& {SU(3)_{C}}\, .
\end{eqnarray}
The groups $G_{TC}$ and $SU(2)_W$ are ordinary technicolor and weak
interactions; the strong and hypercharge groups labeled ``h'' couple
to 3rd-generation fermions and have stronger couplings than the ``$\ell$''
groups coupling to light fermions.  Each fermion has Standard charges
under the groups it couples to.  The strong $U(1)_h$ ensures
that the bottom quark will not condense.
Below the scale $M$, the Lagrangian includes effective interactions
for $\psi_L = (t,b)_L, t_R$, and $b_R$:
\begin{equation} 
{{4\pi \kappa_{tc}}\over{M^2}} \left[\overline{\psi}\gamma_\mu  
{{\lambda^a}\over{2}} \psi \right]^2 -{{4\pi \kappa_1}\over{M^2}} \left[{1\over3}\overline{\psi_L}\gamma_\mu  
\psi_L + {4\over3}\overline{t_R}\gamma_\mu  t_R
-{2\over3}\overline{b_R}\gamma_\mu  b_R
\right]^2 \, . 
\end{equation}
Only the top quark will condense if the interaction strengths satisfy the relationship\,\cite{ehs-topcolor}:
\begin{equation}
\kappa^t = \kappa_{tc} +{1\over3}\kappa_1 >
\kappa_c  >
\kappa_{tc} -{1\over 6}\kappa_1 = \kappa^b\, ,
\end{equation}
where the critical
value is $\kappa_c \approx 3\pi/8$ in the NJL approximation\,\cite{ehs-njl}.

This model combines the strong points of topcolor\,\cite{ehs-topcolor}
and extended technicolor\,\cite{ehs-etc}, giving a more complete
picture of the origin of mass. Technicolor\,\cite{ehs-technicolor}
causes most of the electroweak symmetry breaking; the top condensate
contributes only a small portion.  The $U(1)_h$ charges of the
technifermions can be isospin symmetric to limit contributions to
$\Delta\rho$.  ETC dynamics at a scale $M \gg 1$TeV generates the
light fermion masses and contributes about a GeV to the heavy
fermions' masses.  Finally, the top condensate provides most of $m_t$
and the $t$-$b$ mass splitting.

\section{Phenomenology of Strong Top Dynamics}

Three classes of models of new strong top dynamics with distinctive
spectra are topcolor\,\cite{ehs-topcolor}, flavor-universal extended
color\,\cite{ehs-fucolorons}, and top
seesaw\,\cite{ehs-topseesaw}. Exotic particles in these models include
colored gauge bosons (topgluons, colorons), color-singlet gauge bosons
(Z'), composite scalar states (top-pions, q-pions), and heavy fermions
(usually, but not always\,\cite{ehs-extension}, weak singlets).  A
search for $Z' \to t\bar t$ is addressed in
ref. 13. Searches for composite scalars are closely
related to those for Higgs
bosons\,\cite{ehs-rmoore,ehs-scalars}.  Here, we focus on
new colored gauge bosons and fermions.

\subsection{Topcolor Models}

Topcolor models\,\cite{ehs-topcolor} include an extended $SU(3)_h \times
SU(3)_\ell$ color group and an extended $U(1)_h \times U(1)_\ell$
hypercharge group; in each case, third generation fermions transform
under the stronger ``h'' group and the other fermions, under the
weaker ``$\ell$ '' group (table \ref{tab:topcolor}).  The weak
interactions are Standard.  The low-energy spectra of these models
typically include a heavy color-octet of gauge bosons called topgluons
and a heavy Z' boson, all primarily coupled to the $t$ and $b$
quarks.  The new strong interactions affecting $t$ and $b$ quarks also
form $t\bar t$, $b\bar{b}$ and $t\bar{b}$ composite scalars.

CDF has looked for signs of Z' and topgluons in $b\bar{b}$ final
states\,\cite{ehs-topgluons}.  Their limit on $\sigma\cdot B{X \to
b\bar{b}}$ for a narrow resonance X falls just short of restricting
the mass of a topcolor Z'.  Topgluons, on the other hand, are expected
to be quite broad due to their large coupling to quarks.  CDF sets
limits of 280 (340, 640) $<$ M $<$ 670 (375, 560) GeV on a topgluon
with a width equal to 30\% (50\%, 70\%) of its mass.  More
recently\,\cite{ehs-rmoore} CDF has set a limit on a leptophobic $Z'$
decaying to $t\bar t$; if $\Gamma(Z') = .012 M_{Z'} (.04 M_{Z'})$ the
bound is $M_{Z'} > 480$ GeV (780 GeV).  The Run II and LHC experiments
will extend these searches in both the $b\bar{b}$ and $t\bar{t}$
channels.

\begin{table}[t]
\caption{Fermion strong and hypercharge gauge charges in (a) topcolor
and (b) flavor-universal coloron models. Each row refers to a
different fermion generation. A charge of ``SM'' has the same value as
in the Standard Model.}
\label{tab:topcolor}
\vspace{0.4cm}
\begin{center}

\hspace{-4.5cm}
\begin{minipage}{1in}
\begin{tabular}{|l|c|c|c|c|}
\hline

{\bf (a)}& $\bf SU(3)_h$ & $\bf SU(3)_\ell$ 
& $\bf U(1)_h$ & $\bf U(1)_\ell$ 
\\ \hline 
I   & SM &   1 &   0  &   SM \\
II  &   SM &   1 & 0  &   SM \\
III &   SM &   1 &   SM &   0 \\
\hline
\end{tabular} \end{minipage}\hspace{5.5cm}
\begin{minipage}{1in}\begin{tabular}{|l|c|c|c|c|}
\hline
{\bf (b)}& $\bf SU(3)_h$ & $\bf SU(3)_\ell$ 
& $\bf U(1)_h$ & $\bf U(1)_\ell$ 
\\ \hline 
I   & SM &   1 &   0  &   SM \\
II  &   SM &   1 & 0  &   SM \\
III &   SM &   1 &   SM &   0 \\
\hline
\end{tabular} \end{minipage}

\end{center}
\end{table}

\subsection{Flavor-Universal Coloron Models}

In flavor-universal theories\,\cite{ehs-fucolorons} all quarks
transform as triplets under only the strong $SU(3)_h$ group, as in
table \ref{tab:topcolor}.  Hence, in addition to a Z' boson
preferentially coupled to third-generation fermions, these models
include a color-octet of coloron bosons that couple equally strongly
to all quark flavors.  The coloron coupling to quarks is enhanced
relative to the QCD interaction strength by a factor of $\cot\theta
\equiv g_h / g_\ell$.  The composite scalars in these models thus
include not only the top-pions of topcolor models but also a full
range of ``q-pions'' including quarks of all flavors; this greatly
reduces potential contributions to flavor-changing neutral currents.

\begin{figure}[b]
\centerline{\epsfig{figure=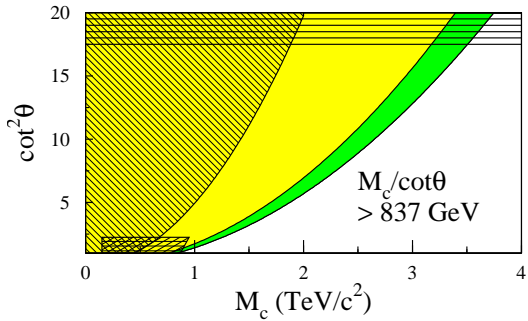,height=1.7in}}
\vspace{-.6cm}
\caption[collim]{Experimental limits\protect{\,\cite{ehs-bertram}} on the mass and
coupling strength of flavor-universal colorons.}
\label{fig:coloron}
\end{figure}

Limits\,\cite{ehs-bertram} on flavor-universal colorons are shown in
figure \ref{fig:coloron}.  Like topgluons, the colorons are generally
quite broad.  Hence, CDF's search for new narrow resonances decaying
to dijets\,\cite{ehs-tev2000} applies only to relatively
weakly-coupling topgluons with small $\cot\theta$ (cross-hatched
region).  Light colorons would make noticeable contributions to
$\Delta\rho$ and are excluded in the diagonally hatched
region\,\cite{ehs-fucolorons}.  The shape of the D\O\ dijet angular
distribution is sensitive to somewhat heavier colorons, and excludes
the light-shaded area\cite{ehs-d0ang}.  Finally, the shape of the D\O\
dijet mass spectrum places a bound\,\cite{ehs-bertram} $M_c/\cot\theta
> 837$ GeV on the coloron mass and coupling (dark-shaded region).  The
horizontally hatched region lies in a different phase of the
model\,\cite{ehs-fucolorons}.

The strongly-coupled colorons that trigger top quark condensation lie
at $\cot^2\theta \approx 17$.  The lower bound on
their mass is $\sim$3.4 TeV, so direct observation of
such states must await the LHC.

\begin{table}[t]
\caption{Third-generation fermion charges in a top seesaw model.} \label{tab:topseesaw}
\vspace{0.4cm}
\begin{center}
\begin{tabular}{|l|c|c|c|}
\hline
 & $\bf SU(3)_h$ & $\bf SU(3)_\ell$ &  $\bf SU(2)$ \\ \hline \hline
$\bf (t,\, b)_L$   &  3 &  1 &  2 \\
$\bf t_R,\ b_R$  &  1 &  3 &  1 \\
$\bf \chi_L$ &  1 &  3 &  1 \\
$\bf \chi_R$ &  3 &  1 &  1 \\
\hline
\end{tabular}
\end{center}
\end{table}

\subsection{Top Seesaw Models}

In top seesaw models\,\cite{ehs-topseesaw}, only the strong interaction
has an extended gauge structure; the weak and hypercharge groups are
as in the Standard Model.  At a minimum\,\cite{ehs-extension}, the third
generation of quarks is augmented by a colored weak-singlet state
$\chi$.  The strong and weak charges of the $t$, $b$ and $\chi$ quarks
are as shown in table \ref{tab:topseesaw}.  Because it is the $t_l$
and $\chi_R$ which are triplets under the strong $SU(3)_h$, the
composite scalar that contributes to electroweak symmetry breaking and
$m_t$ is made of $\bar t_L \chi_R$ rather than $t\bar{t}$ as in
topcolor models.  Hence the top quark acquires mass through a seesaw
mechanism which indirectly links $t_L$ with $t_R$ through the
intervening $\chi$ states:

\vspace{.2cm}

\begin{math}
\hspace{2cm}\hbox{\epsfysize=.3truein \epsfbox{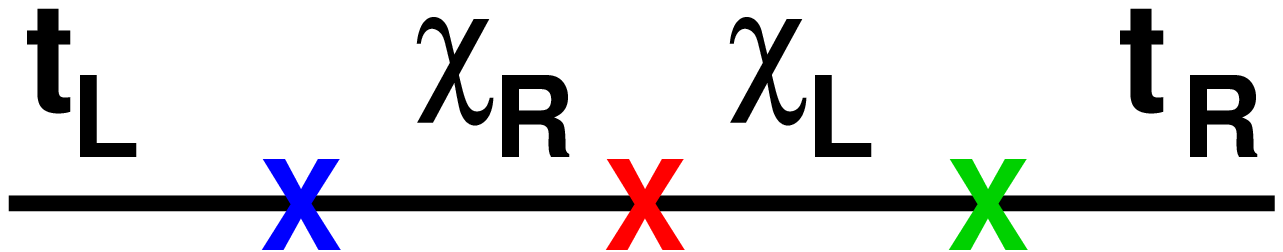}} \hspace{2cm} \raise5pt\hbox{
$ \left( \begin{array}{cc} \bar{t}_L & \bar{\chi}_L \end{array} \right)
\left( \begin{array}{cc} 0 & {m_{t \chi}} \\ 
{\mu_{\chi t}} & {\mu_{\chi \chi}} \end{array} \right) 
\left( \begin{array}{c} t_R \\ \chi_R \end{array} \right)$} 
\end{math}

\vspace{.3cm}

The size of the CDF and D\O\
top dilepton samples\,\cite{ehs-dilepton} provide a lower bound on the masses of
weak-singlet quarks mixing with ordinary quarks.  A heavy mass
eigenstate ($q^H$) whose left-handed component is mostly weak-singlet
could contribute to the dilepton sample via the process 
$ p \bar{p} \to q^H \bar{q}^H \to q^L W \bar{q}^L W 
\to q^L \bar{q}^L \ell \nu_\ell \ell' \nu_{\ell'}$ 
Note that while $q^H$ pairs are produced at the same rate as top quark
pairs of equal mass, the small weak charge of $q^H$ suppresses its
charged-current branching fraction $B(q^H \to q^L W)$.  For $b^H$,
Cabibbo suppression further inhibits charged-current decays and the
flavor-conserving neutral current decay $B(b^H \to b^L Z)$ dominates.
The CDF and D\O\ data imply\,\cite{ehs-singlet} that $M_{d^H,\ s^H} \geq
140$ GeV and that (if all quarks have weak-singlet partners) $M_{b^H}
\geq 160$ GeV.  Since precision electroweak data indicates that the
$\chi$ states have masses of order several TeV in some
models\,\cite{ehs-georgi}, the direct searches are not yet very
restrictive.

\section{Conclusions}

Modern theories of dynamical electroweak symmetry breaking and mass
generation include new strong dynamics peculiar to the top and bottom
quarks.  Hence, these models predict new particles at scales up
to 10 TeV with signatures in jets or heavy flavors: technihadrons,
topgluons, colorons, exotic quarks, Z' bosons, and composite scalars.
Experiments are already searching for these new states, and there
is plenty of room for exciting discoveries at Run II.

\section*{Acknowledgments}
Work supported in part by the U.S. Department of Energy
under grant DE-FG02-91ER40676.

\end{document}